\documentclass[10pt,letterpaper]{article}
\usepackage[top=0.85in,left=1in,footskip=0.75in,marginparwidth=2in]{geometry}

\usepackage[utf8]{inputenc}

\usepackage{cite}
\usepackage{amsmath}
\usepackage{soul}
\usepackage{xcolor}

\usepackage{nameref}
\usepackage[hidelinks]{hyperref}

\usepackage{microtype}
\DisableLigatures[f]{encoding = *, family = * }

\usepackage{changepage}

\usepackage[aboveskip=1pt,labelfont=bf,labelsep=period,singlelinecheck=off]{caption}

\makeatletter
\renewcommand{\@biblabel}[1]{\quad#1.}
\makeatother

\usepackage{lastpage,fancyhdr,graphicx}
\usepackage{epstopdf}
\fancyheadoffset[L]{2.25in}
\fancyfootoffset[L]{2.25in}

\usepackage{color}

\definecolor{Gray}{gray}{.25}

\usepackage{graphicx}

\usepackage{sidecap}

\usepackage{wrapfig}
\usepackage[pscoord]{eso-pic}
\usepackage[fulladjust]{marginnote}
\reversemarginpar

\begin{document}
\vspace*{0.35in}

\begin{center}
{\Large
\textbf\newline{Monte Carlo simulations as a route to compute probabilities}
}
\newline
\\
Parasuraman Swaminathan\textsuperscript{*}
\\
\bigskip
Electronic Materials and Thin Films Lab, \\
Ceramics Technologies Group - Center of Excellence in Materials and \\Manufacturing for Futuristic Mobility, \\
Dept. of Metallurgical and Materials Engineering, \\
Indian Institute of Technology, Madras, Chennai, India
\\

\bigskip
*Email: swamnthn@iitm.ac.in

\end{center}

\section*{Abstract}
Monte Carlo simulations are based on the manipulation of random numbers to evaluate probable outcomes, with applicability in a variety of different fields. By assigning probabilities, which can be determined \textit{a priori}, to various events, it is possible to track the evolution of the system over length and time scales which are not normally accessible to other simulation techniques. Monte Carlo simulations can provide insights, which can be used to develop more realistic models. In this work, these simulations are used to model a variety of probability problems normally encountered at the high school and undergraduate level. The simulations are used to introduce concepts related to system size, simulation runs (repeatability), and basic statistics. While many of the problems discussed here have analytical expressions, systems where easy analytical solutions are not available are also discussed. In this work, MATLAB is used to model these systems and the \textit{rand} function in the software is used extensively for random number generation.    


\section {Introduction}

Monte Carlo simulations depend on random sampling in order to obtain numerical results. The method is different from deterministic approaches, which normally involve solving analytical expressions from basic mathematical and scientific principles. Monte Carlo simulations, on the other hand, allow us to interrogate multiple solution pathways and statistical analysis of these pathways is carried out to predict potential outcomes. Probabilities can be assigned to these outcomes based on their relative frequencies\cite{Samik2008, Bird1981}. The name originates from the Monte Carlo casino in Monaco and these simulation were originally used in modeling neutron diffusion in fissionable materials\cite{Eckhardt1987, Morin2019}. The various pathways can either have equal probabilities or could be weighted depending on \textit{a priori} information, obtained either by experimental observation or by other deterministic routes such as first principles calculations\cite{Yoder1994, Yoshimoto2000} or CALPHAD\cite{De2011}.  

Monte Carlo simulations have been used in a variety of scientific fields. Particularly in the area of materials science they haven been used for modeling diffusion-limited aggregation\cite{Hayakawa1986} (to model nanoparticle growth on surfaces\cite{Antonov2003, Antonov2004}), sputter deposition\cite{Mtohiro1986, Biersack1987}, thin film growth\cite{Zhang2004}, solvent evaporation leading to particle aggregation (coffee-ring effect)\cite{Crivoi2014}, and a variety of other examples. They are also used in other branches of engineering\cite{Billinton1999} and sciences and many times are used jointly with other simulation techniques\cite{Molnar2012, Miller2015}.

In this work, Monte Carlo simulations are used to solve problems in basic probability to illustrate the power and versatility of this approach. Effect of sample size (or number of simulation runs) and extracting of basic statistical data from these simulations are described. Most of the systems described here have simple analytical solutions so that comparison with the results of the simulations are straightforward. An example where an analytical solution cannot be readily obtained is also described. The simulations are implemented using MATLAB, which particularly lends itself for such virtual experiments\cite{Mordechai2011, Krzyzanek2003}. The \textit{rand} function in MATLAB is used to generate the random numbers for the experiments. Data extraction, visualization, and compilation has also been carried out using MATLAB. 

\section {Coin toss}
A simple example for understanding the role of random numbers is the coin toss experiment. The probability of a coin landing on either `heads' or `tails' is $1/2$ or 0.5. One way of understanding this value, is that, if the coin is tossed twice, heads and tails should have each appeared \textit{exactly} once. But simple experimental verification shows that this is not always the case. In reality, what this probability of 0.5 means is that for a sufficiently `large' number of trials, the number of times heads and tails appear tend to be equal. Now, testing this experimentally is tedious, but a virtual experiment can be easily carried out using MATLAB. 

For the coin toss experiment, since the appearance of \textit{heads} or \textit{tails} is equally probable, a random number less than 0.5 can be assigned as tails, while a number greater than 0.5 becomes heads or vice versa. With simple FOR loops and IF conditional statements, it is possible to run this experiment in MATLAB. For obtaining reasonable statistics, the simulations are run from 10-1000 times, at select iterations such as 10, 20, 50, 100, 250, 500, 750, and 1000. The number of times heads is obtained for each of these runs is measured and normalized by the total number of runs. To illustrate the statistical nature of this process, the experiment is repeated five times (at each iteration) and the spread in the data is used to compute the mean and standard deviation for these experiments.   

\begin{figure}[h]
    \centering
    \includegraphics[width=8.3cm]{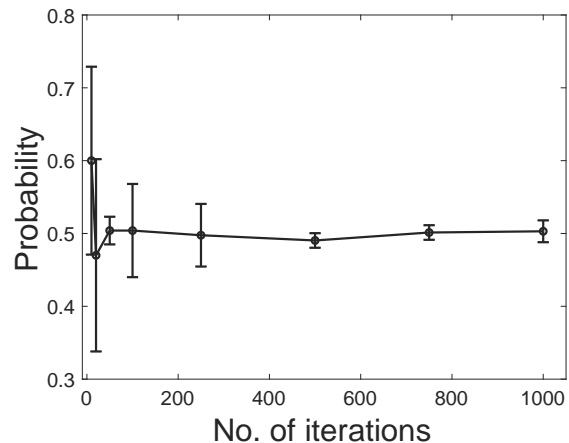}
    \caption{Probability of obtaining heads (tails) in a simple coin toss experiment, as a function of the number of iterations. Five runs were conducted at each step. The standard deviation of the average of each run reduces as the number of iterations increase. The numerical values are tabulated in Table \ref{tab:Table1}.}
    \label{fig:Fig1}
\end{figure}

Consider the data pertaining to a particular iteration (say, $i=10$). The number of times heads (tails) is obtained is calculated for each iteration and normalized by the total number of iteration. This provides the probability of obtaining heads (tails) for that particular run. However, if the same program is executed again the number of time heads (tails) is obtained would be different. To provide some statistics for this process the program is executed five times (at each iteration) to obtain five values of probability. The average and standard deviation of these probabilities, as a function of the number of iterations, are plotted in figure \ref{fig:Fig1} and tabulated in table \ref{tab:Table1}.   

\begin{table}
    \centering
    \caption{Probability of obtaining heads in a simple coin toss experiment, as a function of the number of iterations. Five runs were conducted at each step and the average and standard deviations of these runs are tabulated. The same data is graphically presented in figure \ref{fig:Fig1}.\\}
    \begin{tabular}{|c|c|c|}
    \hline
    \textbf{Iterations} & \textbf{Average} & \textbf{Standard deviation}  \\
    \hline
         10 & 0.600 & 0.129 \\
         20 & 0.470 & 0.132 \\
         50 & 0.504 & 0.019 \\
         100 & 0.504 & 0.064 \\
         250 & 0.498 & 0.043 \\
         500 & 0.490 & 0.010 \\
         750 & 0.501 & 0.010 \\
         1000 & 0.503 & 0.015 \\
         \hline
    \end{tabular}
    \label{tab:Table1}
\end{table}

From the plots it is evident that as the number of iterations increase the averages tend towards the value of 0.5 and more importantly the standard deviation, i.e., the spread in the averages, reduces. The probability of obtaining a head or a tail in a coin toss is 0.5. This means that for an `infinitely' large number of experiments the number of heads and tails would be equal. From a more practical point of view, the data in table \ref{tab:Table1} show that even for 50 iterations (i.e. tossing the coin 50 times) the standard deviation is less than 4 \%. More importantly, compared to 20 iterations, the value drops by nearly an order of magnitude. This means that for 50 iterations, more often than not, we would tend to obtain 25 heads (tails) and increasing the number of iterations beyond that (at least till 1000) does not seem to have any visible impact on either the average or the standard deviation. The coin toss represents a simple system, with only two outcomes of equal probability. As the complexity of the system increases, the number of iterations required to simulate `infinite number of trials' increases. This can be well illustrated using the example of a dice. 

\section {Dice(s) rolling}
\subsection{Single dice}
An experiment similar in nature to coin toss is the rolling of an unbiased dice. A dice has six faces and an unbiased dice has equal probability of landing on any face when rolled. Thus, the problem is similar to coin toss except that the  probability of landing on a particular face (number) is $1/6$ ($0.167$) rather than $1/2$. Thus, the same protocol that was used for the coin toss experiment can be adopted here. 

\begin{figure}[h]
    \centering
    \includegraphics[width=8.3cm]{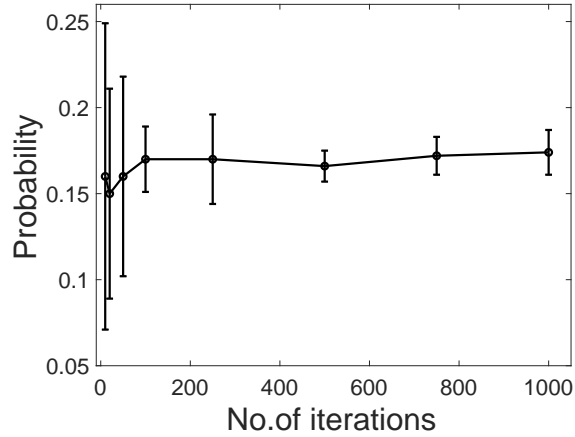}
    \caption{Probability of obtaining a particular number during the rolling of an unbiased dice, as a function of the number of iterations. Five runs were conducted at each step. The standard deviation of the average of each run reduces as the number of iterations increase. The numerical values are tabulated in Table \ref{tab:Table2}.}
    \label{fig:Fig2}
\end{figure}

The data from these experiments are summarized in figure \ref{fig:Fig2} and table \ref{tab:Table2}. A behaviour similar to coin toss is seen with respect to the averages and the standard deviation for the unbiased dice rolling experiments. The standard deviation is large when the iterations are small and reduces as the number of iterations increase. Significantly, a reduction of one order of magnitude in the standard deviation is seen only when the number of iterations reaches 500 and is comparable for 750 and 1000 iterations. Compared to the data from the coin toss (when an order of magnitude difference is seen at 50 iterations) this shows that `infinite number of trials' for this system is larger than the case of the coin toss by an order of magnitude. This difference can be attributed to the reduced probability of each occurrence ($1/6$) when compared to the coin toss experiment ($1/2$). This also shows that as the number of choices increases (reduction in individual probability) the number of experiments required to mimic the theoretical answer also increases. To further illustrate this point, we can look at the rolling of two unbiased dice and the probabilities of obtaining different sums in this distribution.       

\begin{table}[h]
    \centering
    \caption{Probability of obtaining a particular number during the rolling of an unbiased dice, as a function of the number of iterations. Five runs were conducted at each step and the average and standard deviations of these runs are tabulated. The same data is graphically presented in figure \ref{fig:Fig2}.\\}
    \begin{tabular}{|c|c|c|}
    \hline
    \textbf{Iterations} & \textbf{Average} & \textbf{Standard deviation}  \\
    \hline
         10 & 0.160 & 0.089 \\
         20 & 0.150 & 0.061 \\
         50 & 0.160 & 0.058 \\
         100 & 0.170 & 0.019 \\
         250 & 0.170 & 0.026 \\
         500 & 0.166 & 0.009 \\
         750 & 0.172 & 0.011 \\
         1000 & 0.174 & 0.013 \\
         \hline
    \end{tabular}
    \label{tab:Table2}
\end{table}

\subsection{Rolling of two unbiased dice}
An extended system to the rolling of a single dice is the rolling of two unbiased dice (a similar problem can be crafted with coin toss but this system offers more possibilities to explore). In such a case, the sum of the two faces can be between 2-12, with different probabilities for each sum. These values can be arrived at analytically by considering the various combinations of obtaining a particular sum and are presented in table \ref{tab:Table3}. For example, to obtain the sum 3, there are two possibilities, i.e., (2,1) or (1,2), while to obtain the sum 7 there are 6 possibilities, i.e., (6,1), (5,2), (4,3), (3,4), (2, 5), and (1, 6). 

\begin{table}[h]
    \centering
    \caption{Probability of obtaining a sum ranging from 2 to 12 when two unbiased dice are rolled. Each sum has a different probability since the number of possible combinations are different. \\}
    \begin{tabular}{|c|c|}
    \hline
    \textbf{Sum} & \textbf{Probability}  \\
    \hline
         2 & $1/36$ \\
         3 & $2/36$ \\
         4 & $3/36$ \\
         5 & $4/36$ \\
         6 & $5/36$ \\
         7 & $6/36$ \\
         8 & $5/36$ \\
         9 & $4/36$ \\
         10 & $3/36$ \\
         11 &  $2/36$ \\
         12 & $1/36$ \\
         \hline
    \end{tabular}
    \label{tab:Table3}
\end{table}

From the knowledge of these probabilities it is possible to model this experiment using Monte Carlo simulations by directly using the probability to obtain a particular sum (as shown in table \ref{tab:Table3}). However, Monte Carlo simulations also allow us to model this system even if this information is not available. This is because it is possible to individually consider the rolling of each dice, as described in section 3.1, and then use the sum of the faces to obtain the necessary probabilities. As a test case, we can consider two situations, obtaining a sum of 3 and 7 and compare the modeling data with the analytical results. 

There are some differences in the way this experiment is executed in MATLAB when compared to the previous experiments. Since two dice are being rolled, the total number of combinations is 36 (for a single coin toss the number is 2 and for a single dice it is 6). Hence, 10 iterations would not be sufficient to provide the necessary statistical information since it is below the total number of possible combinations. Hence, the minimum number of iterations is fixed at 50. Similarly, the maximum number of iterations is also increased taking into account the possible combinations and thus the other iterations evaluated are 100, 250, 500, 1000, 2500, 5000, and 10000. Each of these is executed 5 times to obtain the average and standard deviation. In order to calculate the sum of the two dice, the value obtained from the \textit{rand} function in MATLAB is converted into an integer between 1 and 6 by using the \textit{ceil} function (ceiling or rounding up to the nearest integer). Two random numbers are generated, one to represent each dice, and are converted to integers before adding them. The number of instances that a particular sum is obtained is then counted and normalized with respect to the total number of iterations to obtain the probability. The technique adopted here can be easily extended to more than two dice (or even two or more coins with suitable modifications). 

\begin{figure}[h]
    \centering
    \includegraphics[width=8.3cm]{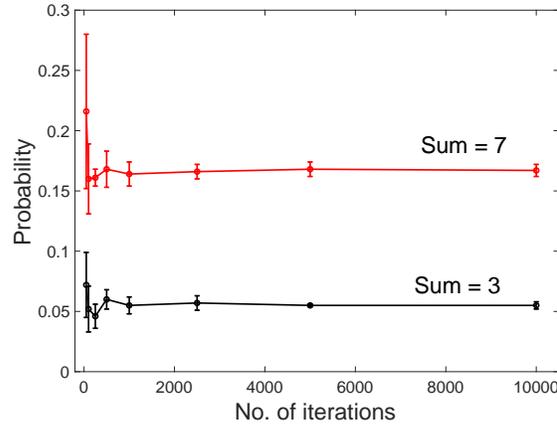}
    \caption{Probability of obtaining a particular sum from the rolling of two unbiased dice. Two values, 3 and 7, are considered. The probability vs. the number of iterations is plotted and five runs were considered for each iteration. It can be seen that the standard deviations are low only for a large number of iterations due to the reduced probability of each event.}
    \label{fig:Fig3}
\end{figure}

The data from the simulations are presented in figure \ref{fig:Fig3}. Two values considered are 3 and 7. The probability for obtaining these values are listed in table \ref{tab:Table3} and are 0.056 and 0.167 respectively. This difference arises because there are more possible combinations to obtain a sum of 7 than 3. As can be seen from the plots, for a small number of iterations there is considerable scatter in both the average probability and the standard deviation. Compared to the case of a single coin and dice, the standard deviation values are reduced only for more than 1000 iterations. There is very little change in the mean and standard deviation beyond 1000 iterations. These data points are consistent with the trend that lower the probability of a particular event more the number of iterations are required to obtain values consistent with analytical results. Thus, as the number of dice are increased beyond two, more trials would be needed in order to obtain statistically reproducible results.   

\section {Random selection}
A particularly interesting set of probability problems include what is described as random selection. A simple example of this would be choosing, for example, a white ball from a bag containing a mixture of white and black balls. In general, if a bag contains $n$ white balls and $m$ black balls, the probability of choosing a white ball is just $n/(n+m)$. This is the case if replacement is allowed, i.e., the ball is replaced in the bag after choosing so that the total number is unchanged. If there is no replacement, then there are $n-1$ white balls and $m$ black balls and the new probability of choosing a white ball is $(n-1)/(n+m-1)$ and so on. These problems can become more complicated and analytical solutions difficult to obtain. However, such problems can also be solved using Monte Carlo simulations. As an illustration, consider the following scenario described below. 

\textit{Consider a bag containing 50 white and 50 black socks. A person randomly draws two socks from the bag, one after the other and in quick succession, and wears them without noticing their color. The process is repeated by 49 other people. Now, if all these 50 people are assembled in a room, what is the probability that 25 of them will be wearing socks of the same color, either white or black? }

It is hard to solve a such a problem analytically because of the large number of combinations that are possible. Also, there are two situations which can be considered, one with replacement and one without replacement. 

\subsection{With replacement}
In the case of replacement each person is presented with a bag with 50 white and 50 black socks from which they can draw either two white or two black socks. An alternate way of stating this is that the replacement happens after each person has drawn two socks. Let us consider a general scenario of a bag having $n$ white socks and $m$ black socks. In general, $n \neq m$. The probability of picking a white sock ($p_w$) or black sock ($p_b$) is given by
\begin{equation}
    \begin{split}
        p_w \: =\: \dfrac{n}{n+m} \\
        p_b \: =\: \dfrac{m}{n+m} \\
        p_w\;+\;p_b \: =\: 1
    \end{split}
\end{equation}
Now, the probability of picking two white socks in a row (replacement happens only after both socks are picked) ($p_{2w}$) is given by
\begin{equation}
    p_{2w}\: =\: \dfrac{n}{n+m} \times \dfrac{n-1}{n+m-1}
\end{equation}
The second term in the above equation arises because after the first white sock is picked, the number remaining is ($n-1$). Similarly, the probability of picking two black socks in a row (without replacement) ($p_{2b}$) is
\begin{equation}
    p_{2b}\: =\: \dfrac{m}{n+m} \times \dfrac{m-1}{n+m-1}
\end{equation}
Thus, for a bag containing 50 white and 50 black socks, the probability of picking two white socks is 0.247, which is the same as the probability exists for picking two black socks. The process will be repeated for each person. However, to obtain the probability that 25 people will we wearing socks of the same colour (either black or white) we need to consider all possible sequences in which 25 people have either black or white socks. Solving the above problem analytically is extremely tedious due to the large number of possible sequences. 

A simpler way to obtain the probabilities is to use Monte Carlo simulations, making use of the probabilities from equation (1) along with a random number generator to decide whether a white or black sock is picked. By repeating this process, a given pair of random numbers can decide whether two white or two black or a white-black (black-white) combination is picked. When this is carried out 50 times (since there are 50 people) we can get the information on how many white or black pairs are picked. This constitutes one trial. To obtain reliable statistics, a large number of trials can be conducted and a histogram of the number of white and black pairs can be obtained. Dividing the counts by the total number of trials provides the probability. Thus, a system which is hard to solve analytically can be solved using a simple Monte Carlo simulation. 

\begin{figure}[h]
    \centering
    \includegraphics[width=8.3cm]{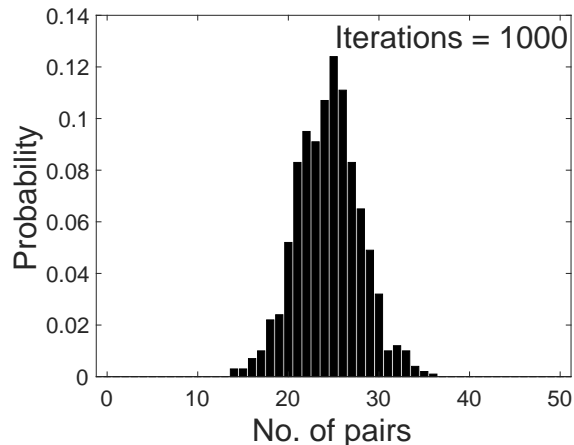}
    \caption{Histogram of the number of pairs (white or black) obtained when two pairs are randomly picked from a bag containing 50 white and black socks, with replacement. There are a total of 50 people who pick a pair and the socks are replaced after each pick. This trial is run 1000 times to obtain reliable statistics. The probability is obtained by normalizing the counts by the total number of iterations.}
    \label{fig:Fig4}
\end{figure}

In MATLAB, the solution can be implemented using a series of FOR loops with IF-ELSE statements. Figure \ref{fig:Fig4} shows the histogram distribution obtained after running 1000 iterations (trials). The data is normalized by the total number of iterations in order to directly obtain the probabilities. From the figure, the maximum probabilities are obtained in the middle, i.e., around to 20-30 pairs of white or black. The probability drops to zero below 10 or above 40. This could be attributed to the fact that only 1000 trials are conducted and because of the low probability of these events they are not accessed by the MC simulation.   

\begin{figure}[h]
    \centering
    \includegraphics[width=8.3cm]{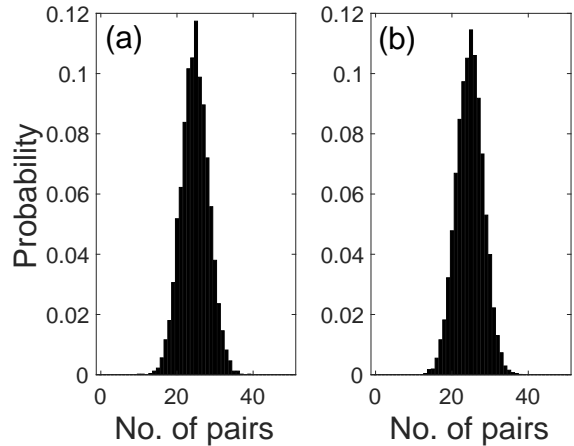}
    \caption{Histogram of the number of pairs (white or black) obtained when two pairs are randomly picked from a bag containing 50 white and black socks, with replacement. There are a total of 50 people who pick a pair and the socks are replaced after each pick. Two trial sizes are considered, (a) 5000 and (b) 10000 times. The probability is obtained by normalizing the counts by the total number of iterations.}
    \label{fig:Fig5}
\end{figure}

In order to test the effect of the number of trials, simulations were performed for 5000 and 10000 trials. The results are plotted in figure \ref{fig:Fig5}. Similar distributions to figure \ref{fig:Fig4} are obtained with the maximum value obtained for 26 pairs. This value obtained was 0.117, for both 5000 and 10000 iterations, and the probability drops sharply as the number of pairs increases or decreases. To obtain the probability of picking 20 or 30 pairs, the data can be directly read from the plots in figures \ref{fig:Fig4} and \ref{fig:Fig5}. Thus, using Monte Carlo simulations, it is possible to access a larger phase space than possible from analytical or numerical calculations.  

\subsection{Without replacement}

The problem becomes considerably trickier when we consider the situation without replacement. Thus, the first person to pick the socks has 50 to choose from, the next person has 48, and so on, until the last person has only two socks to choose from. Given the random nature of the process, at each stage there can exist unequal quantities of white and black socks (except for the first person). Interestingly, without replacement the probability of obtaining 25 people wearing the same colour socks (black or white) actually turns to be zero. To understand consider the fact that we start out with 100 socks in total (50 black and 50 white). Let us assume that 25 people have the same color socks (black or white) and in a particular case there are 12 people wearing white socks and 13 people wearing black socks. Hence, 24 white and 26 black socks are accounted for among them leaving, 26 white and 24 black socks for distribution among the rest. However, it is not possible to distribute this such that everyone will have one white and one black sock, since one pair of white socks will remain. Hence, when we consider the situation without replacement it is only possible to have even pairs of all black or all white and odd pairs will be disallowed. This situation is also borne out during the Monte Carlo simulation. This is a modified version of the program executed in the previous section but disallowing for replacement.

\begin{figure}[h]
    \centering
    \includegraphics[width=8.3cm]{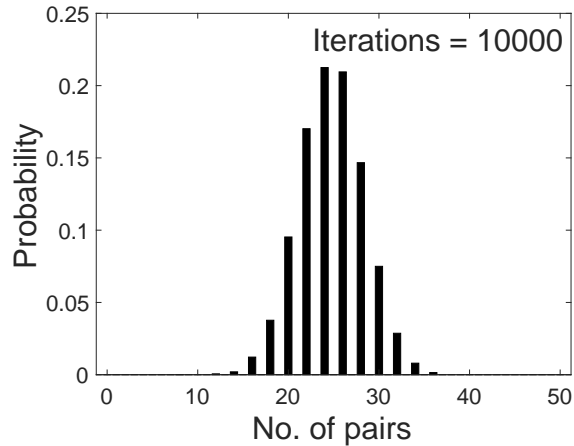}
    \caption{Histogram of the number of pairs (white or black) obtained when two pairs are randomly picked from a bag containing 50 white and black socks, without replacement. There are a total of 50 people who pick a pair one after the other until the bag is empty. This trial is run 10000 times to obtain reliable statistics. The probability is obtained by normalizing the counts by the total number of iterations. Only even combinations are allowed since it not possible to have odd number of pairs under this present condition.}
    \label{fig:Fig6}
\end{figure}

The results of the simulation, in the form of a probability plot, are summarized in figure \ref{fig:Fig6}. This plot is obtained after performing 10000 iterations to collect reliable statistics, which is more than what was carried out in the previous situation with replacement. As can be seen from the plot, only even combinations are allowed since it is not possible to have odd pairs of all white or black socks, when we start out with equal number of both colours. In the probability plot in figure \ref{fig:Fig6} all odd pairs have probability zero, which is contrasted to the data shown in figures \ref{fig:Fig4} and \ref{fig:Fig5}, which is for the case with replacement. The maximum probabilities are obtained from either 24 or 26 pairs, which is expected. The probability values obtained here are higher than the case described earlier since the number of outcomes are smaller. We can roughly see that the probabilities are nearly twice of the case with replacement since the number of available outcomes (only even values) is half. The values will change if we start with unequal number of black and white socks and these variations can be interrogated easily using Monte Carlo simulations.     

\newpage

There are however some limitations to using Monte Carlo method to solve probability problems. One issue relates to the simulation size (defined as the number of iterations required). Probability defines the average outcome for an `infinitely' large number of trials. Since, it is practically impossible to conduct infinite trials the solution would be to always run the smallest number of trials beyond which the outcomes does not change appreciably. For very simple systems, which are computationally inexpensive, the number of trials do not matter. All the examples considered in this manuscript are small and hence large trials can be easily run to obtain `accurate' results. The same need not be true for other systems. The number of iterations needed is related to the probabilities of the individual events. A simple rule of thumb is that rarer an event, more number of trials are required to capture the event. For systems such as a single coin or an unbiased dice, the individually events are equally probable (either heads or tails or any one number from 1-6). Similarly, for two dice the spread in probability values are small. Hence, the individual probabilities can be captured by choosing a suitable number of iterations. However, for the system described in section 4, where obtaining number of pairs below 10 or above 40 have very low probabilities, a large number of simulations need to be carried out. Most of the time the results would not lie in the region of interest and hence obtaining such low probability events are computationally exhaustive. Monte Carlo simulations are more suited for situations where analytical solutions are hard and probabilities can be reasonably estimated by running a series of trial. More the number of trials, greater will be the confidence in the probabilities estimated from these simulations.  

\section{Conclusion}

Monte Carlo simulations are a useful tool for analyzing situations and systems where both temporal and spatial scales are not accessible to other traditional techniques. By setting up `virtual experiments`, with either \textit{a priori} known or estimated frequencies (rates), it is possible to access a large phase space of the system to estimate probabilities. These simulations are particularly well-suited for systems where analytical calculations are hard but the underlying system can be easily modeled using standard programming. In this manuscript, Monte Carlo simulations have been used to calculate probabilities for a series of commonly discussed situations. While analytical results exist for coin toss and dice rolling, an example of random pairing is used to illustrate a system where simple analytical solutions do not exist. The effect of the number of trials on the estimated probabilities is described and it was shown that more trials are needed when the underlying probability values are low. Monte Carlo simulations also cannot access events with relatively low probabilities compared to other outcomes and special techniques are needed to access these situations. Overall, they offer a convenient way to estimate probabilities in a variety of situations. The systems that can be tackled depend on the existence of \textit{a priori} rates and the skill of the programmer(s).

\section*{Acknowledgments}
This manuscript was prepared under the aegis of the course Computational Materials Engineering Lab (MM3110), part of the undergraduate curriculum at the Department of Metallurgical and Materials Engineering at IIT Madras. The feedback of the students of the July-August 2021 batch is appreciated, especially with respect to the problem outlined in section 4. The author also acknowledges the support of his colleagues, Dr. Anand Krishna Kanjarla and Dr. Satyesh Kumar Yadav, who are also co-instructors of this course. Support from the Ceramics Technologies Group - Center of Excellence in Materials and Manufacturing for Futuristic Mobility (project number SB/2021/0850/MM/MHRD/008275) is also acknowledged. 


\bibliography{library}

\bibliographystyle{unsrt}

\end{document}